\documentclass[12pt]{article}

\usepackage{amsmath,amsfonts,amssymb,amsthm} \usepackage{color, graphicx} \usepackage{epsfig,fullpage} \usepackage{natbib} \usepackage{url} \usepackage{lineno} \usepackage{subfigure} \usepackage{mathtools,MnSymbol}
\usepackage{soul}
\usepackage{url}
\usepackage{fullpage} 
\usepackage[utf8]{inputenc} 
\usepackage{kotex}
\usepackage{amssymb}
\usepackage{verbatim}
\usepackage{lscape}
\usepackage{tabularx}
\usepackage{multirow}
\usepackage{adjustbox}
\usepackage{geometry}
\usepackage{color}
\usepackage{natbib}
\usepackage[blocks]{authblk}  
\usepackage[justification=centering]{caption} 
\usepackage{enumerate}
\usepackage{array}
\usepackage[normalem]{ulem} 
\usepackage{threeparttable} 
\usepackage{stackrel}
\usepackage{hyperref}
\hypersetup{colorlinks=true, citecolor=blue}

\newtheorem{theorem}{Theorem}

\begin{document}
\title{Testing for Stochastic Order in Interval-Valued Data}

\renewcommand\Affilfont{\itshape\small}

\author[1]{Hyejeong Choi}

\author[1]{Johan Lim}
\author[2]{Minjung Kwak}

\author[1]{Seongoh Park\footnote{To whom all correspondence should be addressed. Email: \texttt{inmybrain@snu.ac.kr}}}

\affil[1]{Department of Statistics, Seoul National University, Seoul, Korea} 
\affil[2]{Department of Statistics, Yeungnam University, Gyeongsan, Korea}

\date{}

\maketitle 

\begin{abstract}  
\noindent 
We construct a procedure to test the stochastic order of two samples of interval-valued data. We propose a test statistic which belongs to U-statistic and derive its asymptotic distribution under the null hypothesis. We compare the performance of the newly proposed method with the existing one-sided bivariate Kolmogorov-Smirnov  test using real data and simulated data.
\end{abstract}
\baselineskip 18pt 

\section{Introduction}

\hskip5mm We discuss the two-sample tests for stochastic order of two interval-valued samples. In the interval-valued data, the variable of interest is not observed as a single point but is displayed in the form of an interval, with lower and upper bounds. 
For example, interval-valued data is observed when stock price is reported  monthly by lower and upper limit prices. In addition, blood pressure data, which motivates our research, has diastolic blood pressure (DBP) and systolic blood pressure (SBP) as lower and upper bounds.
It is a fundamental problem in statistics to test the stochastic order of two populations as well as to verify the equality of the two distributions. However, little research has been done for the interval-valued data; even definition of stochastic order for interval-valued data is not clearly established. Thus, this paper introduces its definition and proposes a method to test the stochastic order of two samples of interval-valued data. 

The remainder of the paper is organized as follows. In section 2, we define the stochastic order of interval-valued data. In section 3, we propose a test statistic for testing the order of interval-valued data and derive its asymptotic null distribution using the general theory on U-statistic. In section 4, we examine the performance of the modified two-dimensional Komogorov-Smirov(K-S) statistic and the proposed through a numerical study. In section 5, we apply the methods to the blood pressure data from female students in the US. In section 6, we conclude the paper with a summary.

\section{Simple stochastic order}
	
Before we introduce the notion of the stochastic order for interval-valued data, we look at the stochastic order for the usual univariate case. Let $X$ and $Y$ be two univariate random variables such that
$$
\text{Pr}(X > z) \le \text{Pr}(Y>z),\quad \text{for all } z \in \mathbb{R}.
$$
Then, $Y$ is said to be \textit{stochastically greater than} $X$ (denoted by $X \le_{st} Y $). If additionally $\text{Pr}(X > z)< \text{Pr}(Y>z)$ for some $z$, then $Y$ is said to be \textit{stochastically strictly greater than} $X$ \citep{Shaked:2006}.

The stochastic order for interval-valued data can be defined similarly. Let $\mathbf{x}=(\ell_1, u_1]$ and $\mathbf{y}=(\ell_2, u_2]$ be two intervals. Then we denote $\mathbf{x} < \mathbf{y} $ and say $\mathbf{y}$ is \textit{greater} than $\mathbf{x}$ if $ \ell_1 <  \ell_2$ and $u_1 < u_2$.
Now, let $\mathbf{X}$ and $\mathbf{Y}$ be two random intervals such that
\begin{equation}\label{eqn:st1}
	\text{Pr}(\mathbf{X} >\mathbf{z}) \le \text{Pr}(\mathbf{Y} > \mathbf{z}),\quad \text{for all interval } \mathbf{z}.
\end{equation}
Then, $\mathbf{Y}$ is said to be \textit{stochastically greater than} $\mathbf{X}$ and denoted by $\mathbf{X} \le_{st} \mathbf{Y} $.
Let $\overline{F}(\mathbf{x})=\text{Pr}(\mathbf{X} >\mathbf{x})$ and $\overline{G}(\mathbf{y})=\text{Pr}(\mathbf{Y} >\mathbf{y})$ be the survival functions of the random intervals $\mathbf{X}$ and $\mathbf{Y}$, respectively. Then, (\ref{eqn:st1}) is equivalent to
\begin{equation}\nonumber 
\overline{F}(\ell, u)  \le \overline{G} (\ell, u)~~~ \text{for all } (\ell, u): \ell < u.
\end{equation}
\begin{figure}[!tbp]
	\begin{center}
		\includegraphics[width=8cm]{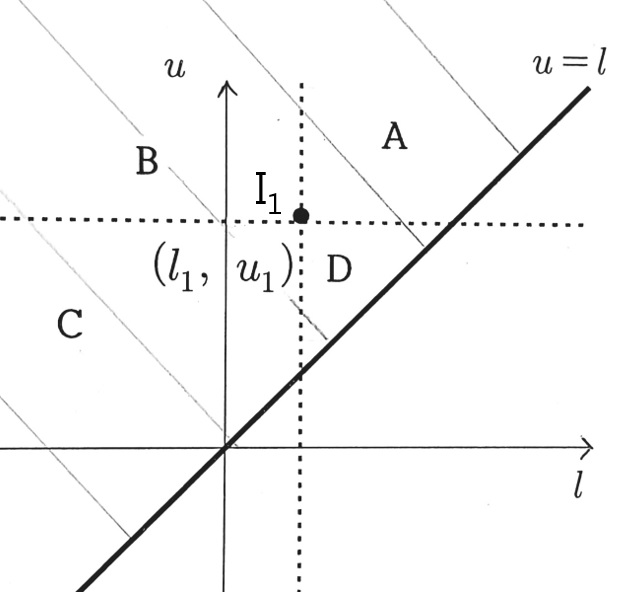}
		\caption{\small A graphical illustration of the order of interval-valued data. Region A, B, C, and D are respectively defined by intersection of the half-plane $u>\ell$ with the first, second, third, and fourth quadrant when $I_1=(\ell_1, u_1]$ is set as the origin.}
		\label{fig: interval_plane}	
	\end{center}
\end{figure}
We can illustrate the order of the intervals as follows (see Figure \ref{fig: interval_plane}). Let the interval $(\ell_1, u_1]$ denoted by the point $(\ell_1, u_1)$ in the plane. Note that in the plane, interval-valued data is displayed at the top of the line $ u = \ell $  due to the constraint $\ell < u$.  Any interval-valued data of the half-plane belongs to any of three cases according to the order relation with the interval $I_1=(\ell_1, u_1]$.
\begin{enumerate}
	\item region A: intervals are greater than $I_1=(\ell_1, u_1]$.
	\item region C: intervals are less than $I_1=(\ell_1, u_1]$.
	\item region B or D: intervals do not have an order relation with  $I_1=(\ell_1, u_1)]$.
\end{enumerate}
For the last case, 
an interval $(\ell_B, u_B]$ in region B satisfies $(\ell_1, u_1] \subset (\ell_B, u_B]$, while an interval $(\ell_D, u_D]$ in region D satisfies $(\ell_D, u_D] \subset (\ell_1, u_1]$.


\section{Test statistic}
Let us consider two independent samples of random intervals. Suppose that a first sample $\mathbf{X}_i=(\ell_{1i}, u_{1i}]$, $i=1, \ldots, m$, has a survival function $\bar{F}$ and the second sample $\mathbf{Y}_j=(\ell_{2j}, u_{2j}]$, $j=1, \ldots, n$, has a survival function $\bar{G}$.
We want to verify the null hypothesis that both samples come from an identical distribution, ``$\mathcal{H}_0$: $\bar{F} (\mathbf{z})= \bar{G}(\mathbf{z})$  for all $\mathbf{z}$'' against to the alternative hypothesis 
that $\mathbf{Y}$ is stochastically strictly greater than $\mathbf{X}$, i.e., 
``$\overline{F}(\mathbf{z}) \le \overline{G} (\mathbf{z})$ for all $\mathbf{z}$ and $\overline{F}(\mathbf{z}) < \overline{G} (\mathbf{z})$ for some interval $\mathbf{z}$''.

The statistic we propose to test the stochastic order is 
\begin{equation} \label{eqn:test-stat} 
T= \frac{1}{mn} \sum_{i=1}^m\sum_{j=1}^n S_{i j},
\end{equation}
where 
\begin{equation} \nonumber 
S_{ij} =
\left\{ \begin{array}{rl}
~1  ~~&\text{if} ~~ 	\ell_{1i}<\ell_{2j} \text{ and } u_{1i}<u_{2j},\\
~-1  ~~&\text{if}~~  	\ell_{1i}>\ell_{2j} \text{ and } u_{1i}>u_{2j},\\
~0 ~~&\text{otherwise.} 
\end{array}
\right.
\end{equation} 
Note that under the null $\bar{F}=\bar{G}$, $\text{Pr}(S_{ij}=1)=\text{Pr}(S_{ij}=-1)$ and thus $\mathbb{E}(T)=0$.

The statistic $T$ belongs to a class of U-statistics, which allows one to derive its asymptotic null distribution based on the asymptotic theory of the U-statistic. We introduce below a general asymptotic theory of U-statistics reported in Chapter 6 of \citet{Lehmann:1999}.  
Let $\phi(x_1, \ldots, x_{a} ; y_1, \ldots, y_{b})$ be a symmetric  kernel of $a+b$ ($1 \le a \le m, 1 \le b \le n$) arguments. Here, the symmetric kernel denotes a function whose value does not change by changing the order of arguments $(x_1, \ldots, x_{a})$ or $(y_1, \ldots, y_{b})$. Let $\theta$ defined below be a parameter of interest;
$$
\theta=\theta(\bar{F}, \bar{G}) = \mathbb{E} [\phi(\mathbf{X}_1, \ldots, \mathbf{X}_{a} ; \mathbf{Y}_1, \ldots, \mathbf{Y}_{b})],
$$
and define its U-statistic by
\begin{equation}\label{eqn:ustat_two}
U_{m,n} = \binom{m}{a}^{-1} \binom{n}{b}^{-1}\sum_{C_{m, a}} \sum_{C_{n, b}} \phi(\mathbf{X}_{i_1}, \ldots, \mathbf{X}_{i_{a}} ; \mathbf{Y}_{j_1}, \ldots, \mathbf{Y}_{j_{b}}),
\end{equation}
where $C_{k,t}$ is the collection of all subsets of $\{1,2,\ldots, k\}$ with size $t$ and dummy indices running over summations are $(i_1, \ldots, i_a)$ and $(j_1, \ldots, j_b)$, respectively. $U_{m,n}$ is an unbiased estimator of $\theta$ and its variance is
\begin{equation} \nonumber 
{\rm Var}(U_{m,n}) =\sum_{i=1}^a\sum_{j=1}^b \frac{\binom{a}{i}\binom{m-a}{a-i}}{\binom{m}{a}} \frac{\binom{b}{j}\binom{n-b}{b-j}}{\binom{n}{b}} \sigma_{ij}^2,
\end{equation} 
where $\sigma_{ij}^2$ is given by
\begin{eqnarray} 
\sigma_{ij}^2 &=& {\rm Cov}[\phi(\mathbf{X}_1, \dots, \mathbf{X}_i, \mathbf{X}_{i+1}, \ldots, \mathbf{X}_a; \mathbf{Y}_1, \ldots, \mathbf{Y}_j, \mathbf{Y}_{j+1}, \ldots, \mathbf{Y}_b), \nonumber \\
&& \qquad \qquad \phi(\mathbf{X}_1, \dots, \mathbf{X}_i, \mathbf{X}_{i+1}^{\prime}, \ldots, \mathbf{X}_a^{\prime}; \mathbf{Y}_1, \ldots, \mathbf{Y}_j, \mathbf{Y}_{j+1}^{\prime}, \ldots, \mathbf{Y}_b^{\prime})], \nonumber
\end{eqnarray}
and $\mathbf{X}_i^{\prime}$ and $\mathbf{Y}_j^{\prime}$ are independent copies of $\mathbf{X}_i
$ and $\mathbf{Y}_j$. 
The theorem below from Chapter 6 of \citet{Lehmann:1999} explains the asymptotic distribution of the U-statistic (\ref{eqn:ustat_two}) above.
\begin{theorem}[Lehmann(1999), Theorem 6.1.3 (ii)]\label{theo:ustat}
		As $m/N \rightarrow \rho \in (0, 1)$ and $N=(m+n) \rightarrow \infty$, $\sqrt{N} \big(U_{m,n}-\theta \big)$ converges in distribution to the normal distribution with mean $0$ 
	and variance $\sigma^2= \frac{a^2}{\rho}\sigma_{10}^2+\frac{b^2}{1-\rho}\sigma_{01}^2$. Here, $\sigma_{10}^2$ and $\sigma_{01}^2$ are computed by
\begin{eqnarray} 
	\sigma_{10}^2&=& {\rm Cov}[\phi(\mathbf{X}_1, \mathbf{X}_2, \ldots, \mathbf{X}_a;\; \mathbf{Y}_1, \ldots, \mathbf{Y}_b),\;  \phi(\mathbf{X}_1, \mathbf{X}_2^{\prime}, \ldots, \mathbf{X}_a^{\prime};\; \mathbf{Y}_1^{\prime}, \ldots ,\mathbf{Y}_b^{\prime})] \in (0,\infty),  \nonumber\\
\sigma_{01}^2 &=& {\rm Cov}[\phi(\mathbf{X}_1, \ldots, \mathbf{X}_a;\; \mathbf{Y}_1,\mathbf{Y}_2 \ldots,\mathbf{Y}_b),\;  \phi(\mathbf{X}_1^{\prime}, \ldots, \mathbf{X}_a^{\prime};\; \mathbf{Y}_1, \mathbf{Y}_2^{\prime}, \ldots, \mathbf{Y}_b^{\prime})] \in (0,\infty). \nonumber 
	\end{eqnarray} 
\end{theorem}
\noindent 
Applying the general theory above for U-statistics to our case, we can derive the asymptotic null distribution of our $T$ statistic.
\begin{theorem} \label{theo:ustat2} 
	Under the null hypothesis that $\mathcal{H}_0: \bar{F}=\bar{G}$, if $m/N \rightarrow \rho \in (0, 1)$ as $N=(m+n) \rightarrow \infty$, then
	\begin{equation}\nonumber 
\sqrt{N} T ~\xrightarrow{d} ~{N} \left(\,0\,, \,  \frac{ \theta_1 + \theta_2 -2 \theta_3}{\rho(1-\rho)} \,\right),
\end{equation} 
where  $\theta_1 = {\rm Pr}\big(\mathbf{X} < \min(\mathbf{Y}, \mathbf{Y}^\prime)\big)$, $\theta_2 = {\rm Pr}\big(\max(\mathbf{Y}, \mathbf{Y}^\prime)  < \mathbf{X} \big)$, and $\theta_3 = {\rm Pr}\big(\mathbf{Y}^\prime  < \mathbf{X} < \mathbf{Y}\big)$.
\end{theorem}
\noindent
Parameters $\theta_1,\theta_2, \theta_3$ used to compute the asymptotic variance can be approximated by permuting observations within each sample. 
To understand it, we observe the followings.
\begin{equation}\label{eq:apprx_theta1}
\begin{array}{rcl}
\theta_1 &=& {\rm Pr}\big(\mathbf{X} < \min(\mathbf{Y}, \mathbf{Y}^\prime)|F=G\big) \\[0.5em]
&=&  {\rm Pr}\big(\mathbf{X} < \min(\mathbf{X}^\prime, \mathbf{X}^{\prime\prime})|F=G\big)\\[0.5em]
&=& {\rm Pr}\big(\mathbf{X} < \min(\mathbf{X}^\prime, \mathbf{X}^{\prime\prime})\big) \Big(={\rm Pr}\big(\mathbf{Y} < \min(\mathbf{Y}^\prime, \mathbf{Y}^{\prime\prime})\big) \Big),
\end{array}
\end{equation}
where $\mathbf{X}, \mathbf{X}^\prime, \mathbf{X}^{\prime\prime}$ are independent random intervals from the first population. Consequently, $\theta_1$ can be approximated by
$$
\hat{\theta}_1 = \dfrac{\sum_{i,j,k: \text{distinct}} {\rm I}\big(\mathbf{X}_i < \min(\mathbf{X}_j, \mathbf{X}_k)\big)}{2m(m-1)(m-2)}
+ \dfrac{\sum_{i,j,k: \text{distinct}} {\rm I}\big(\mathbf{Y}_i < \min(\mathbf{Y}_j, \mathbf{Y}_k)\big)}{2n(n-1)(n-2)}.
$$
Equation (\ref{eq:apprx_theta1}) has an implication that the above approximation would be a valid estimate of $\theta_1$ even under the alternative hypothesis. 

\begin{proof}
	For $\mathbf{x}=(\ell_{1}, u_{1}], \mathbf{y}=(\ell_{2}, u_{2}]$, let us define 
	$\phi(\mathbf{x}; \mathbf{y})= {\rm I}(\mathbf{x} < \mathbf{y}) - {\rm I}(\mathbf{x} > \mathbf{y}) = {\rm I}(\ell_{1}<\ell_{2} ,  u_{1}<u_{2})-{\rm I}(\ell_{1}>\ell_{2} ,  u_{1}>u_{2})$.
	Then, $T$ can be presented by a two sample U-statistic when  $a=b=1$;
	\begin{equation}\nonumber 
U_{m,n}=\frac{1}{mn} \sum_{i=1}^m\sum_{j=1}^n \phi(\mathbf{X}_i; \mathbf{Y}_j).
\end{equation}
Therefore, by applying Theorem \ref{theo:ustat}, we have 
	\begin{equation} \nonumber 
\sqrt{N} (U_{m,n}-\theta) ~\xrightarrow{d} ~{N} \left(\,0\,, \, \frac{\sigma_{10}^2}{\rho}+\frac{\sigma_{01}^2}{1-\rho}\,\right),
\end{equation} 
where $\theta= \mathbb{E}(\phi(\mathbf{X}; \mathbf{Y}))$, $\rho= \lim \frac{m}{N} \in (0, 1)$, $\sigma_{10}^2=
{\rm Cov}\left[ \phi(\mathbf{X}; \mathbf{Y}), \phi(\mathbf{X}; \mathbf{Y}^{\prime})\right]$, and $\sigma_{01}^2= {\rm Cov}\left[ \phi(\mathbf{X}; \mathbf{Y}), \phi(\mathbf{X}^{\prime}; \mathbf{Y})\right]$.
	
	Now, let us denote interval random variables by $\mathbf{X}=(L_{1}, U_{1}]$, $\mathbf{Y}=(L_{2}, U_{2}]$, and $\mathbf{Y}^{\prime}=(L_{2}^{\prime}, U_{2}^{\prime}]$.
	Under the null hypothesis $\bar{F}=\bar{G}$, we have $\theta= \ \mathbb{E}(\phi(\mathbf{X}; \mathbf{Y}))=\text{Pr}(L_{1}<L_{2}, U_{1}<U_{2})-\text{Pr}(L_{1}>L_{2}, U_{1}>U_{2}) =0$. The variance component  $\sigma_{10}^2$ (= $\sigma_{01}^2$) is evaluated as 
	\begin{align*} 
	\sigma_{10}^2 &= {\rm Cov}\left[ \phi(\mathbf{X}; \mathbf{Y}), \phi(\mathbf{X}; \mathbf{Y}^{\prime})\right] \\
	&=  \mathbb{E}\left[\phi(\mathbf{X}; \mathbf{Y}) \phi(\mathbf{X}; \mathbf{Y}^{\prime}) \right] \quad (\because \theta =0 ) \\
	&= \mathbb{E}[{\rm I}(\mathbf{X} < \mathbf{Y}) {\rm I}(\mathbf{X} < \mathbf{Y^\prime})] - \mathbb{E}[{\rm I}(\mathbf{X} < \mathbf{Y}) {\rm I}(\mathbf{X} > \mathbf{Y^\prime})]\\
	& \qquad - \mathbb{E}[{\rm I}(\mathbf{X} > \mathbf{Y}) {\rm I}(\mathbf{X} < \mathbf{Y^\prime})] + \mathbb{E}[{\rm I}(\mathbf{X} > \mathbf{Y}) {\rm I}(\mathbf{X} > \mathbf{Y^\prime})]\\
	& = \text{Pr}\big(\mathbf{X} < \min(\mathbf{Y}, \mathbf{Y}^\prime)\big) - \text{Pr}\big(\mathbf{Y}^\prime  < \mathbf{X} < \mathbf{Y}\big)\\
	& \qquad - \text{Pr}\big(\mathbf{Y}< \mathbf{X} <  \mathbf{Y}^\prime \big)
	+ \text{Pr}\big(\max(\mathbf{Y}, \mathbf{Y}^\prime)  < \mathbf{X} \big).
	\end{align*}
	Now, we write $\theta_1 = \text{Pr}\big(\mathbf{X} < \min(\mathbf{Y}, \mathbf{Y}^\prime)\big)$, $\theta_2 = \text{Pr}\big(\max(\mathbf{Y}, \mathbf{Y}^\prime)  < \mathbf{X} \big)$, and $\theta_3 = \text{Pr}\big(\mathbf{Y}^\prime  < \mathbf{X} < \mathbf{Y}\big)$. Thus, under $\bar{F}=\bar{G}$, we get
	\begin{equation} \nonumber 
	\sigma_{10}^2=\sigma_{01}^2 = \theta_1 + \theta_2 -2\theta_3.
	\end{equation} 
	Hence, the asymptotic variance of $\sqrt{N}T(=\sqrt{N}U_{m,n})$ is 
\begin{equation} \nonumber 
\frac{\sigma_{10}^2}{\rho}+\frac{\sigma_{01}^2}{1-\rho}= \frac{ \theta_1 + \theta_2 -2 \theta_3}{\rho(1-\rho)}.
\end{equation} 
\end{proof}

\section{Numerical study}

In this section, we compare the power of our proposed test (denoted as ``U-test'') to one-sided bivariate K-S test (denoted as ``K-S test'').
U-test can be classified by how its null distribution is approximated. ``U-{\rm perm}'' designates U-test where we approximate the null distribution by a permutation method, while ``U-{\rm asym}'' is the one depending on the approximation given in Theorem \ref{theo:ustat2}. K-S test for the alternative hypothesis $\overline{F} < \overline{G}$ is given by \citep{Feller:1948}
\begin{equation}\nonumber 
D_{m,n}^{+}=  \left( \frac{mn}{m+n} \right)^{1/2} \sup_{s, t \in \mathbb{R},\; s<t} \left(\widehat{F}_{m} (s, t)-\widehat{G}_{n} (s, t) \right),
\end{equation}
where 	
$\widehat{F}_{m} (s, t) = \frac{1}{m} \sum_{i=1}^m {\rm I} (L_{1i} \le s, U_{1i} \le t)$ and $\widehat{G}_{n} (s, t) =$ $ \frac{1}{n} \sum_{j=1}^n {\rm I} (L_{2j} \le s, U_{2j} \le t)$.
The null distribution of $D_{m,n}^{+}$ is approximated using a permutation method \citep{Gail:1976}.

In the study, to generate interval-valued data $(L, U]$, we consider a transformation to obtain $C=(L+U)/2$ and half-range $R=(U-L)/2$. 
We consider two underlying distributions for $(C, \log R)$; bivariate normal distribution and  bivariate $t$ distribution with the degrees of freedom $5$. 
\begin{eqnarray}
N \left( \begin{pmatrix}
\mu_{\scriptscriptstyle \rm C} \\ \mu_{\scriptscriptstyle \rm R} \end{pmatrix}\;, \begin{pmatrix}
1 & \rho \\ \rho & 1 \end{pmatrix} \right) 
\quad \text{or} \quad 
t_5
\left(
\begin{pmatrix}
\mu_{\scriptscriptstyle \rm C} \\ \mu_{\scriptscriptstyle \rm R}
\end{pmatrix}, 
\begin{pmatrix}
1 & \rho \\ \rho & 1 
\end{pmatrix} 
\right)  \nonumber
\end{eqnarray}
For two populations, we consider $\Pi_1$ and $\Pi_2$ parameterized as follows;
\begin{eqnarray} 
\Pi_1 :&& \mu_1= (0,0), ~~ \Sigma_1= \begin{pmatrix}
1 & \rho \\ \rho & 1 \end{pmatrix}  \nonumber  \\
\Pi_2 :&& \mu_2= (\delta, 0), ~~ \Sigma_2=\Sigma_1. \nonumber
\end{eqnarray}
\begin{figure}[!tbp]
	\begin{center}
		\includegraphics[width=10cm]{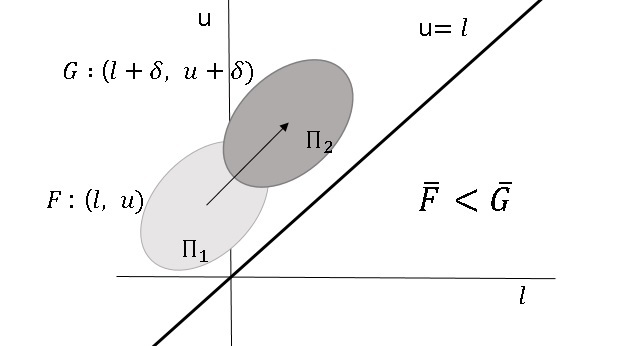}	
		\caption{ \small A graphical illustration of two populations in the simulation study}
		\label{fig: interval_order}	
	\end{center}
\end{figure}
\noindent
For $\delta$, the following four values are used : $(0, ~0.3, ~0.5, ~1.0)$ where $\delta >0$ indicates the alternative hypothesis.
Figure \ref{fig: interval_order} shows the graphical illustration of the simulation setting.
To examine the effect of correlation between the center and range, we use three values for $\rho=(0,~0.4, ~0.8)$.  
The significance level $\alpha$ is set as $0.05$. The size and power are evaluated as the rejection rate among $2,000$ replicates. The number of permutations to generate a null distribution is set as $20,000$. For the sample size $(m, n)$, we consider following 4 cases: (30, 30), (30, 120), (50, 50), (50, 200).

\begin{table}[!tb]
	\caption{\small Simulation results for the stochastic order tests. The power of each test is displayed. The first column denotes the distribution of $(C, \log R)$ : ``N'' indicates the normal case and ``T'' indicates $t-$distribution with df $5$. ``U-perm'' and ``U-asym'' represent two types of U-tests, where the former approximates the null distribution by a permutation method and the latter by the asymptotic result in Theorem 4.2. B-KS denotes the bivariate K-S test.}
	\centering 
	\begin{adjustbox}{max width=\linewidth}
		\begin{tabular}{c|c|c|*3c|*3c|*3c} 
			\hline
			\multirow{2}{*}{case}&\multirow{2}{*}{$(m, n)$} &\multirow{2}{*}{$\delta$} & \multicolumn{3}{c|}{$\rho=0$} & \multicolumn{3}{c|}{$\rho=0.4$} & \multicolumn{3}{c}{$\rho=0.8$} \\
			\cline{4-12} 
			\multirow{2}{*}{}& \multirow{2}{*}{} & \multirow{2}{*}{} & U-perm & U-asym  & B-KS & U-perm & U-asym  & B-KS & U-perm & U-asym  & B-KS \\
			\hline\hline 
			\multirow{16}{*}{(N)}& \multirow{4}{*}{$(30, 30)$}	&	0.0	&	0.045	&	0.044	&	0.042	&	0.043	&	0.042	&	0.041	&	0.045	&	0.045	&	0.040	\\
			\multirow{16}{*}{} &\multirow{4}{*}{}	&	0.3	&	0.301	&	0.293	&	0.158	&	0.307	&	0.306	&	0.178	&	0.425	&	0.425	&	0.289	\\
			\multirow{16}{*}{} &\multirow{4}{*}{}	&	0.5	&	0.573	&	0.568	&	0.321	&	0.599	&	0.598	&	0.366	&	0.789	&	0.788	&	0.630	\\
			\multirow{16}{*}{} &\multirow{4}{*}{}	&	1.0	&	0.980	&	0.979	&	0.829	&	0.988	&	0.988	&	0.900	&	0.999	&	0.999	&	0.995	\\
			\cline{2-12}
			\multirow{16}{*}{}& \multirow{4}{*}{$(30, 120)$}	&	0.0	&	0.049	&	0.047	&	0.052	&	0.051	&	0.050	&	0.058	&	0.050	&	0.050   &	0.046	\\
			\multirow{16}{*}{} &\multirow{4}{*}{}	&	0.3	&	0.396	&	0.393	&	0.267	&	0.422	&	0.420	&	0.312	&	0.578	&	0.578	&	0.489	\\
			\multirow{16}{*}{} &\multirow{4}{*}{}	&	0.5	&	0.745	&	0.744	&	0.551	&	0.781	&	0.780	&	0.619	&	0.929	&	0.928	&	0.876	\\
			\multirow{16}{*}{} &\multirow{4}{*}{}	&	1.0	&	0.999	&	0.999	&	0.979	&	1.000	&	1.000	&	0.991	&	1.000	&	1.000	&	1.000	\\
			\cline{2-12}
			\multirow{16}{*}{}& \multirow{4}{*}{$(50, 50)$}	&	0.0	&	0.055	&	0.055	&	0.042	&	0.054	&	0.054	&	0.040	&	0.049	&	0.049	&	0.040	\\
			\multirow{16}{*}{} &\multirow{4}{*}{}	&	0.3	&	0.411	&	0.412	&	0.252	&	0.436	&	0.439	&	0.287	&	0.589	&	0.590	&	0.476	\\
			\multirow{16}{*}{} &\multirow{4}{*}{}	&	0.5	&	0.756	&	0.757	&	0.525	&	0.790	&	0.792	&	0.605	&	0.936	&	0.937	&	0.873	\\
			\multirow{16}{*}{} &\multirow{4}{*}{}	&	1.0	&	0.999	&	0.999	&	0.973	&	1.000	&	1.000	&	0.992	&	1.000	&	1.000	&	1.000	\\
			\cline{2-12}
			\multirow{16}{*}{}& \multirow{4}{*}{$(50, 200)$}	&	0.0	&	0.052	&	0.051	&	0.040	&	0.052	&	0.050	&	0.047	&	0.057	&	0.056	&	0.048	\\
			\multirow{16}{*}{} &\multirow{4}{*}{}	&	0.3	&	0.557	&	0.556	&	0.378	&	0.602	&	0.590	&	0.462	&	0.775	&	0.776	&	0.709	\\
			\multirow{16}{*}{} &\multirow{4}{*}{}	&	0.5	&	0.904	&	0.903	&	0.733	&	0.925	&	0.922	&	0.831	&	0.987	&	0.987	&	0.975	\\
			\multirow{16}{*}{} &\multirow{4}{*}{}	&	1.0	&	1.000	&	1.000	&	0.999	&	1.000	&	1.000	&	1.000	&	1.000	&	1.000	&	1.000	\\
			\hline\hline 
			\multirow{16}{*}{(T)}& \multirow{4}{*}{$(30, 30)$}	&	0.0	&	0.055	&	0.052	&	0.042	&	0.050	&	0.050	&	0.045	&	0.050	&	0.042	&	0.047	\\
			\multirow{16}{*}{} &\multirow{4}{*}{}	&	0.3	&	0.239	&	0.238	&	0.171	&	0.253	&	0.265	&	0.188	&	0.334	&	0.354	&	0.271	\\
			\multirow{16}{*}{} &\multirow{4}{*}{}	&	0.5	&	0.467	&	0.488	&	0.302	&	0.491	&	0.518	&	0.346	&	0.663	&	0.664	&	0.542	\\
			\multirow{16}{*}{} &\multirow{4}{*}{}	&	1.0	&	0.934	&	0.936	&	0.752	&	0.949	&	0.952	&	0.810	&	0.991	&	0.991	&	0.919	\\
			\cline{2-12}
			\multirow{16}{*}{}& \multirow{4}{*}{$(30, 120)$}	&	0.0	&	0.052	&	0.048	&	0.053	&	0.051	&	0.048	&	0.048	&	0.053	&	0.044	&	0.046	\\
			\multirow{16}{*}{} &\multirow{4}{*}{}	&	0.3	&	0.349	&	0.324	&	0.225	&	0.370	&	0.351	&	0.246	&	0.479	&	0.476	&	0.344	\\
			\multirow{16}{*}{} &\multirow{4}{*}{}	&	0.5	&	0.650	&	0.634	&	0.419	&	0.685	&	0.676	&	0.446	&	0.843	&	0.844	&	0.632	\\
			\multirow{16}{*}{} &\multirow{4}{*}{}	&	1.0	&	0.987	&	0.988	&	0.817	&	0.993	&	0.992	&	0.849	&	1.000	&	1.000	&	0.867	\\
			\cline{2-12}
			\multirow{16}{*}{}& \multirow{4}{*}{$(50, 50)$}	&	0.0	&	0.053	&	0.052	&	0.044	&	0.049	&	0.050	&	0.044	&	0.046	&	0.044	&	0.040	\\
			\multirow{16}{*}{} &\multirow{4}{*}{}	&	0.3	&	0.350	&	0.333	&	0.215	&	0.367	&	0.362	&	0.246	&	0.490	&	0.486	&	0.361	\\
			\multirow{16}{*}{} &\multirow{4}{*}{}	&	0.5	&	0.661	&	0.650	&	0.426	&	0.686	&	0.691	&	0.490	&	0.852	&	0.849	&	0.687	\\
			\multirow{16}{*}{} &\multirow{4}{*}{}	&	1.0	&	0.993	&	0.993	&	0.852	&	0.996	&	0.998	&	0.881	&	1.000	&	1.000	&	0.893	\\
			\cline{2-12}
			\multirow{16}{*}{}& \multirow{4}{*}{$(50, 200)$}	&	0.0	&	0.050	&	0.054	&	0.052	&	0.049	&	0.048	&	0.051	&	0.051	&	0.046	&	0.056	\\
			\multirow{16}{*}{} &\multirow{4}{*}{}	&	0.3	&	0.482	&	0.494	&	0.278	&	0.499	&	0.494	&	0.306	&	0.690	&	0.644	&	0.453	\\
			\multirow{16}{*}{} &\multirow{4}{*}{}	&	0.5	&	0.845	&	0.860	&	0.517	&	0.863	&	0.860	&	0.563	&	0.965	&	0.958	&	0.708	\\
			\multirow{16}{*}{} &\multirow{4}{*}{}	&	1.0	&	1.000	&	1.000	&	0.825	&	1.000	&	1.000	&	0.834	&	1.000	&	1.000	&	0.843	\\
			\hline\hline	    
		\end{tabular}		 	
	\end{adjustbox}
	\label{table: result_order} 
\end{table}

Table \ref{table: result_order} shows some interesting findings with regard to the proposed U-test. First, the power of our U-test is higher than the one-sided K-S test in all cases under consideration regardless of the magnitude of $\rho$.
Also, it is noted that when it comes to U-test, the powers based on a permutation method and asymptotic results are almost same in all cases, which proves the asymptotic result and its accuracy.
Third, the greater the correlation between center and range, the higher power of each test we can get. This phenomenon can be explained using the Mahalanobis distance between two mean vectors from the null and the alternative. The distance is $(\delta, 0){\begin{pmatrix} 1 & \rho \\ \rho & 1 \end{pmatrix}}^{-1} (\delta, 0) = \delta^2 /(1-\rho^2)$, which is increasing in terms of $\rho$. Specifically, when $\rho$ is $0$, $0.4$, and $0.8$, the corresponding distance is $\delta^2$, $1.2\delta^2$ and $2.8\delta^2$, respectively.

\section{Data example}

In this section, we apply the stochastic order tests to a real dataset. The data we use is obtained from National Heart, Lung, and Blood Institute Growth and Health Study (NGHS), which is a $10$-year cohort study to evaluate the temporal trends of cardiovascular risk factors, such as systolic and diastolic blood pressures (SBP, DBP) based on annual visits of 2,379 African-American and Caucasian girls. The blood pressure (BP) data, which is measured at two levels, can be an example of the MM-type interval-valued data. In this analysis, we only use BP measurements at the first visit and remove subjects with missing values. After all, the total number of subjects is $N=2,256$, where Caucasians and African-American girls are $m= 1,112$ and $n=1,144$, respectively. The goal of this application is to test a hypothesis ``BP of African-American is stochastically greater than that of Caucasian girls''.
\begin{table}[!htbp]
	\caption{\small Descriptive statistics of the BP data by race. Mean and standard deviation (in parenthesis) are given. ``{\rm p-value}'' at the last column is the p-value of t-test on the alternative hypothesis ``the BP of African-American is higher than that of Caucasian''.}
	\centering
	\begin{tabular}{c|*2c|*2c} 
		\hline\hline	
		& Caucasian  & African-American& p-value & \\
		\hline
		mid-BP &  78.67 (9.09)  & 80.13 (8.03)  &  $< 0.001$  \\
		DBP &  56.72 (12.19)  & 58.03 (11.72)   &  $0.005$\\
		SBP & 100.62 (9.28) &  102.23 (8.65) & $< 0.001$\\
		half-range & 21.95(5.89) & 22.10 (6.44)  & $0.279$ \\
		\hline\hline
	\end{tabular} 
	\label{table: race_ori}
\end{table}
\noindent
Table \ref{table: race_ori} shows that SBP, DBP, and their center are significantly higher in African-American than in Caucasian.
Meanwhile, it is confirmed that there is no difference in the range between two groups. These results are very similar to the setting of the numerical study, where centers of two groups are similar, but ranges are different.


Now, we verify whether the BP of African-American is stochastically greater than that of Caucasian based on interval-valued data, instead of marginal distributions. Table \ref{table: BP_order} presents test results of previously compared methods. In all tests, the p-values are smaller than 0.001, which ensures that the BP of African-American is stochastically greater than that of Caucasians.
\begin{table}[!htbp]
	\caption{\small Two-sample order tests for the BP data. ``U-perm'' and ``U-asym'' are U-tests depending, respectively, on a permutation method and the asymptotic result in Theorem 4.2. B-KS denotes the bivariate K-S test.}
	\centering
	\begin{tabular}{c|*3c} 
		\hline\hline
		& U-perm & U-asym  & B-KS \\ 
		\hline
		p-value& $<0.001$ & $<0.001$ & $<0.001$  \\
		\hline\hline
	\end{tabular} 
	\label{table: BP_order}
\end{table}

\section{Conclusion}
In this paper, we introduce the notion of stochastic order between two samples of interval-valued data and propose a test statistic based on U-statistic. We compute the asymptotic null distribution of the proposed statistic. The numerical study shows that the asymptotic distribution approximates the null distribution with accuracy, even with small size of samples. Also, the proposed test has higher power than the one-sided bivariate KS test in all cases we consider. Therefore, it can be said that the procedure proposed in this paper is of great use for testing the order of interval-valued data.

\section*{Notes}
Authors want to inform that this manuscript is an English version of the article written in Korean and 
accepted
at \textit{The Korean Journal of Applied Statistics}.

\section*{Acknowledgements}
We would like to show our gratitude to two anonymous reviewers and the editor of \textit{The Korean Journal of Applied Statistics} for their detailed and instructive comments.

\end{document}